\documentclass{elsarticle}

\usepackage{amssymb,amsmath,graphicx,url}
\usepackage[utf8]{inputenc}
\usepackage{listings}
\usepackage{tabularx}
\usepackage{natbib}
\usepackage{lscape}
\usepackage{graphicx}

\begin{document}

\title{Publication venue recommendation using profiles based on clustering}

\author{Luis M. de Campos \corref{cor1}}
\ead{lci@decsai.ugr.es}
\author{Juan M. Fern\'andez-Luna}
\ead{jmfluna@decsai.ugr.es}
\author{Juan F. Huete}
\ead{jhg@decsai.ugr.es}

\cortext[cor1]{Corresponding author. Tel.: +34958243199;}
\address{Departamento de Ciencias de la Computaci\'on e Inteligencia Artificial, ETSI Inform\'atica y de Telecomunicaci\'on, CITIC-UGR, Universidad de Granada, 18071, Granada, Spain}

\begin{abstract}
In this paper we study the venue recommendation problem in order to help   researchers to identify a journal or conference to submit a given paper. A common approach to tackle this problem is to build profiles defining the scope of each venue. Then, these profiles are compared against the target paper. In our approach we will study how clustering techniques can be used to construct topic-based profiles and use an Information Retrieval based approach to obtain the final recommendations. Additionally, we will explore  how the use of authorship, representing a complementary piece of information, helps to improve the recommendations. 
\end{abstract}


\maketitle

\section{Introduction}

The problem of journal or conference recommendation (venue recommendation in general) tries to identify appropriate publication venues to submit a paper.
It is a difficult problem for several reasons: there is an enormous quantity of possible publication venues; even within a single specific research domain there are thousands of publications; for example, only within computer science, \cite{Wang18} reported in 2018 9,185 conferences and 4,152 journals. Therefore, it is not easy for researchers to be aware of all academic venues that fit their domain of interests. The problem is worse due to the increasing number of papers which contain multidisciplinary research, to the dynamic change in the scope of some venues and also for new (unexperienced) researchers or for (experienced) researchers who are going to move to a new research area.

Selecting a good conference or journal in which to publish a new article is very important in the academic world. Almost every researcher of every institution in every country may be interested in a tool for recommending appropriate publication venues for their research papers. This means, among other things, that the recommended venue should have a good match with the topics discussed in the paper, which may avoid early rejection. Probably a good match between the recommended venue and the research profiles of the authors' paper is also desirable.

This problem can be tackled using the so-called recommender systems \cite{Bobadilla13,Lu15,Portugal18,Zhang19}, which in general recommend items to users based on characteristics of the items and users preferences. Within the academic world, these items can be papers to read (to stay up to date in a given topic) \cite{Beel16,Haruna20}, papers to cite (to include as references) \cite{Huang12,Farber20}, researchers (either possible collaborators, outstanding researchers within a topic or possible referees) \cite{Chaiwanarom15,Cifariello19,Pradhan21,Protasiewicz16} or publication venues, as in our case.

There are a number of systems available on the web that tackle the problem of publication venue recommendation (most of them focused on scientific journals), although detailed information about their characteristics and how they work is not available, except in some cases. Among these systems we can find Elsevier JournalFinder\footnote{http://journalfinder.elsevier.com} \cite{Kang15}, Springer Nature Journal Suggester\footnote{https://journalsuggester.springer.com}, Wiley Journal Finder\footnote{https://journalfinder.wiley.com}, Clarivate Manuscript Matcher\footnote{http://endnote.com/product-details/manuscript-matcher}, Journal/Author Name Estimator\footnote{http://jane.biosemantics.org} \cite{Schuemie08}, Edanz Journal Selector\footnote{https://en-author-services.edanzgroup.com/journal-selector} and JournalGuide\footnote{https://www.journalguide.com}. This shows the great interest that this problem arouses in the scientific community and the publishing world.

The two basic approaches to deal with the venue recommendation problem are based on either the textual content of the paper to be published (as well as the content of the papers already published in the venues) \cite{Errami07,Kang15,Medvet14,Rollins17,Schuemie08,Wang18,Yang12} or on some kind of (scientific social) network, using cites, references, co-authorship... \cite{Kucuktunc13,Luong12,Pham11}. There are also works combining both approaches \cite{Pradhan20,Silva15}.
In this paper we are going to focus essentially on content-based methods, although we will also consider a hybrid method which also uses a very simple network taking into account the venues where the authors of the target paper have already published in the past.

Our proposal is to use clustering techniques applied to the papers published in a set of venues to try to detect the main topics which each of the venues deals with. The found clusters corresponding to each venue will be then used to build thematic (sub)profiles, based on the terms appearing in the papers within the clusters which can be associated with this venue, thus obtaining a distributed representation of the themes which are covered. We hope that a better and explicit understanding of the subjects associated to each venue, together with the terms or words used within these subjects, will imply a better choice of the appropriate publication venue for a target paper.
The experiments carried out to assess the merits of our proposal and to compare it with several baseline methods shall use a collection of journal papers (containing 309,551 papers and 1,002 journals) obtained from PMSC-UGR \cite{Albusac18}, which is a document collection extracted from PubMed and Scopus.

The remaining of the paper is organized as follows: Section \ref{related} examines related work on publication venue recommendation; Section \ref{core} shows the details of our proposal to build subprofiles for publication venues based on clustering methods; Section \ref{expe} includes the experimental setting and an analysis of the results obtained by our methods and the baselines; finally, Section \ref{conclu} outlines our conclusions and future lines of research.

\section{Related Work}
\label{related}

There are some works which use content-based recommendation to tackle the problem of venue recommendation but following a machine learning style \cite{Rollins17,Wang18}. In the first case, binary SVM classifiers are used, one for each venue, whereas in the second case a single multiclass sofmax classifier is considered. Other works also consider content-based recommendation in several ways, but using an information retrieval (IR) style \cite{Errami07,Kang15,Medvet14,Safa18,Schuemie08,Yang12}. In \cite{Schuemie08} the MoreLikeThis algorithm of Lucene is used. \cite{Safa18} uses directly cosine similarity. In \cite{Kang15} the authors use noun phrases (instead of isolated terms or n-grams) in combination with Okapi BM25. Terms are used in \cite{Errami07}, but a sentence alignment algorithm is also employed to refine the initial ranking. In \cite{Medvet14} several methods are proposed: some uses n-gram profiles of each venue, whereas other uses the topics generated by LDA together with a clustering algorithm. LDA topics instead of terms are also considered in \cite{Yang12}, together with stylometrics features.
There are some other works which do not rely on content analysis but on social network analysis. In \cite{Luong12} a coauthorship network is built starting from the authors of the target paper. A coauthorship network is also used in \cite{Pham11} in combination with a clustering algorithm, in order to determine the neighborhood to be used by a collaborative filtering algorithm. Another social network, in this case based on refererences/citations (starting from the references of the target paper) is used by a random walk with restart algorithm in \cite{Kucuktunc13}.
A combination of content-based recommendation (in IR style) with social network analysis (co-authorship and/or references-cites) can also be found in the literature \cite{Pradhan20,Safa18,Silva15}.

There are some works where the goal is also to recommend venues but not with the objective of selecting the best venue to publish a given paper. In some cases the goal is to recommend venues to a researcher which are similar to a given venue \cite{Lu09,vanVeller15} or similar to a given collection of articles \cite{Alhoori17}. In the first case a log analysis of search sessions and an analysis of co-citations between journals are used, whereas in the second case classical collaborative filtering is employed (where users are researchers, items are journals and the ratings are computed taking into account the number of articles of a venue added to the user collection). In other cases there is no explicit input (neither articles nor venues) \cite{Garcia17,Beierle16,Boukhris14,Klamma09}. In all these cases a combination of social network (either co-authorship, citation, or co-participation) analysis with collaborative filtering is used. Another works of this kind (requiring neither articles nor venues as input) are \cite{Chen15,Yu18}, where a random walk with restart is used on a network having co-authorship and author-venue relations.

Most of the methods for recommending the appropriate venues for publishing a target article which use content analysis are based, in a way or another, on one of these two ideas: (1) to join all the articles published in a venue in a single macro document, as a unified representation (a profile) of the corresponding venue, or (2) to keep the articles from a venue as separate entities, thus obtaining a distributed representation of each venue. In the first case the unified representation of each venue is compared with the corresponding representation of the target article, using some similarity measure. In the second case the representation of the target article is also compared with those of all the individual articles in the collection. Then, the corresponding scores of the articles associated to the same venue are combined in some way (using some aggregation measure) to get a score for this venue.

Examples of the first approach can be found in \cite{Silva15} (using terms and key phrases as features), \cite{Medvet14} (using n-grams) and \cite{Yang12} (which uses topics generated by LDA). Examples of the second approach can be seen in \cite{Kang15} (which employs noun phrases as features and average as the aggregation measure), \cite{Rollins17,Luong12} (both using terms and a KNN-based aggregation), \cite{Yang12} (using LDA and average), \cite{Pradhan20} (using LDA and doc2vec), \cite{Schuemie08} (using terms and average) and \cite{Errami07} (employing terms and sum). Even some method based on network analysis instead of content analysis, as \cite{Kucuktunc13} is based on the same idea: They compute, from a citation graph using random walk with restart and starting from the references of the target paper, the scores for a set of related papers, and aggregate the scores of the papers from the same venue by summing them.

\section{Venue subprofiles based on clustering}
\label{core}

As mentioned in the previous section, in order to perform the recommendation two different approaches can be found: On the one hand, considering a single profile/representation of each venue  \cite{Silva15,Medvet14,Yang12} (aggregating all the articles in this venue) and, on the other hand,  considering a completely distributed representation of each venue (with as many subprofiles as papers in this venue) as done for example in \cite{Kang15,Rollins17,Luong12,Pradhan20}.  In both cases, the target article is matched with respect to the venues' (sub)profiles in order to recommend the most relevant venues.

The basic idea of our proposal is to use an intermediate situation between these two extremes. We believe that, by obtaining for each venue a set of homogeneous (in a thematic sense) (sub)profiles, we can better capture the different topics considered within this venue and hence to obtain a better representation of it. We hope that this better representation of each venue will translate into better results for the recommendation venue task. The generation of the subprofiles for each venue will be based on clustering methods. In \cite{deCampos20} and references therein, some applications of clustering methods to build compound profiles can be found. It should be noticed that we are going to obtain a clustering of articles, not a clustering of venues.

Let us consider a set of $n$ publication venues $V=\{v_1,v_2,\ldots,v_n\}$ and a collection of $m$ articles published in these venues $A=\{a_1,a_2,\ldots,a_m\}$, where $v(a_i)=v_j$ means that paper $a_i$ is published in venue $v_j$. Let $A_j$ be the set of articles published in venue $v_j$, i.e. $A_j=\{a_i\in A \,|\, v(a_i)=v_j\}$, with $m_j=|A_j|$.

Each article $a_i$ in $A$, in addition to being associated to a venue, $v(a_i)$, has other characteristics: it contains text; let the ``document'' $d(a_i)$ denote the textual content of $a_i$. Depending on the collection, the available text may be either the title and the abstract (and perhaps keywords) of the article or includes also the full text\footnote{in our experiments in Section \ref{expe} we will use a collection having only title, abstract and keywords of each article.}. We will perform document clustering using the terms appearing in $d(a_i)$ as the features (bag of words approach). An article $a_i$ has also a list of authors, $au(a_i)$. A modification of our recommender system based on clustering will also use the list of authors of the target paper to try to improve the offered recommendations. A paper $a_i$ may have also a list of bibliographic references, although in this work we will not use them.

We are going to apply clustering in a global way, i.e. it will be applied once to the collection of all the articles in $A$. Thus, each obtained cluster will be composed of articles possibly belonging to different venues.
Another option, that we will not consider in this work\footnote{but we intend to explore it in the future.} is to apply clustering in a local way, that is to say to apply a separate clustering only to the articles of each venue $A_j$ (i.e. instead of a single clustering for a large collection of $m$ papers, $n$ different clustering, each for a collection of $m_j$ papers).
As our clustering is global, in order to extract subprofiles for each venue we must decompose each cluster into a set of (sub)clusters, one for each venue having articles in it. Let $K$ be the fixed number of clusters\footnote{this is an input parameter of the cluster algorithm.} and $Cl_k$, $k=1,\ldots,K$ be each one of the obtained clusters of articles in $A$\footnote{notice that we are going to use hard clustering, i.e. $\cup_{k=1}^K Cl_k=A$ and $Cl_r\cap Cl_s=\emptyset$.}. Then $Cl_{jk}=Cl_k\cap A_j$, $k=1,\ldots,K$ is the set of clusters associated to venue $v_j$\footnote{notice also that $\cup_{k=1}^K Cl_{jk}=A_j$ and $Cl_{jr}\cap Cl_{js}=\emptyset$.}. The number of clusters associated to $v_j$ may be strictly lesser than $K$ in case that some sets $Cl_{jk}$ become empty (i.e. some cluster $Cl_k$ does not contain any article from venue $v_j$).

For each of the (at most) $K$ clusters associated to a venue $v_j$ we are going to build a (sub)profile: the text $d(a_i)$ of all the articles $a_i$ that belong to the same cluster $Cl_{jk}$ (which are expected to have a topical similarity) is concatenated to build a single macro-document, $d_{jk}$. Then all these new documents are indexed for use by an information retrieval system (IRS). When the system has to recommend an appropriate venue for a given target article $ta$, the text in this article, $d(ta)$, is used as a query to the IRS, which will then return a ranking of documents/subprofiles. 

As a ranking of venues is the required output, it is necessary to apply a final fusion step to the obtained ranking of subprofiles, in order to aggregate the scores of all the subprofiles which are associated to the same venue and after to rerank the venues. For this purpose, in this paper we propose the use of   $CombLgDCS$ \cite{deCampos17} which, in brief, is the sum of the scores of all the subprofiles associated to the same venue but it also takes into account their positions in the ranking (logarithmic devaluation). Them, the recommended venues will be obtained after sorting in decreasing order of the combined score, $CombScore_t$, defined as:

 \[
 CombScore_t(v_j)  = score_t(d_{mj}) + \sum_{k\in\{1,\ldots, K\}, d_{jk}\neq d_{mj}} \frac{score_t(d_{jk})}{\log_2(pos_{jk}+1)},
 \]

\noindent where $score_t$ is the original score returned by the IRS for the text query,  $d_{mj} = \arg\max_k score_t(d_{jk})$ represents the macro-document associated to $v_j$ with maximum score and $pos_{jk}$ represents the position of the document $d_{jk}$ in the original ranking. Note that this is a hybrid between pure score-based and pure rank-based fusion approaches \cite{Macdonald08}.

\subsection{Using authors as input for venue recommendation}

In this section we are going to explore a different source of evidence for the venue recommendation task, namely the authors of the published papers.
We believe that those venues where a researcher has previously published may also be appropriate venues for future papers. The intuition is that researchers tend to publish in a reduced number of venues, because usually they do not drastically change of research areas, and there is also some kind of effect ``better the devil you know''.
Although our goal is the same, i.e. selecting the appropriate venue where a paper can be published, in this case we will consider a different set of features: the authors of the papers instead of their terms (authorship instead of textual content).

As we already mentioned in Section \ref{related}, there exist several papers in the literature which employ authors information (e.g. \cite{Klamma09, Luong12,Safa18}) in order to find the most promising venues for a target paper. In general, these models build an authors social network, which in the end reflects collaboration between researchers considering co-authorship, citation or co-participation, among others kind of relationships. Then, these networks are finally used as the base of the recommendations using social networks based criteria as PageRank or graph-based similarity \cite{Klamma09}. 

Nevertheless, our proposal in this paper concerning the use of authors is quite different. We still are going to represent venues by means of topical (sub)profiles obtained also through a clustering process of the articles in the collection based on their {\em textual} content\footnote{the two extreme cases being again grouping together all the articles published in the same venue in a single profile, i.e. no clustering at all, and considering a separate subprofile for each article in the venue, i.e. each article forms its own cluster.}. However, instead of building each (sub)profile of a venue as the concatenation of all the terms of the articles appearing in the associated cluster that are published in that venue, we only consider the authors of these articles as the features, i.e. each subprofile is formed only by the authors' names of these articles (which can be repeated). 
So, we do not create an authors' network but we group together authors which work in the same topics\footnote{notice that strictly speaking we are not creating a cluster of authors, we cluster articles about the same topics but represent these articles only through their authors.}.

From the IR perspective, for each subprofile of a venue $v_j$ which is associated to a topic/cluster $Cl_{jk}$, the list of authors $au(a_i)$ of all the articles $a_i$ that belong to $Cl_{jk}$ are concatenated to build a ``document'', $r_{jk}$ (whose corresponding ``terms'' are the authors's names of the articles). The frequency of an author in such a document $r_{jk}$ represents the number of papers published in $v_j$ by such author under the given topic\footnote{binary codification instead of frequency has been also explored.}.

Once we have built these documents $r_{jk}$, they can be indexed by an IRS and given a target paper $ta$, its authors' list $au(ta)$ will be used as a query in order to retrieve the most relevant subprofiles. Note that, when using authors' subprofiles, the three main characteristics used by an IRS to compute the relevance  scores (term frequency $tf$, inverse document frequency $idf$ and document lenght $dl$) are: the number of publications of the author under a topic in a venue, related to $tf$, the specificity of the author (more weight is given to authors that concentrate their publications in a topic or venue), related to $idf$, and the importance of the topic to the venue (large authors subprofiles would correspond to those more common topics in the venue whereas the smaller ones could be associated to more specific or marginal topics), related to $dl$.

As in the previous case of using the textual content instead of the authors to form the documents to be indexed, to obtain a final ranking of venues the $CombLgDCS$ fusion method is used and the combined score, $CombScore_a$, is computed as:

\[
CombScore_a(v_j)  = score_a(r_{mj}) + \sum_{k\in\{1,\ldots, K\}, r_{jk}\neq r_{mj}} \frac{score_a(r_{jk})}{\log_2(pos_{jk}+1)},
\]

\noindent where now $score_a$ represents the score returned by the IRS for an author-based query, $r_{mj} = \arg\max_k score_a(r_{jk})$ and $pos_{jk}$ represents the position of the document $r_{jk}$ in the original ranking.

\subsection{Combining both sources of information} \label{combination}

The fact that the main topics of the target paper have already appeared in other papers published in a given venue, $v_j$, and the fact that its authors have previously published in $v_j$ might be considered two complementary sources of information about the appropriateness of recommending $v_j$. In this case, it is natural to think about a model able to combine them in order to improve the recommendations.

A simple approach could be to consider a linear combination between them. Nevertheless, it must be considered that the set of values (the range) of $CombScore_t$ and $CombScore_a$ may differ considerably. This is because, generally, the IR models do not normalize the output scores and their values depend on the number of terms in the query matching with the documents. So, since the number of matching terms for $CombScore_t$, those in the textual body, should be larger than the number of  authors used to compute $CombScore_a$, we can expect that $CombScore_t$ is much greater than $CombScore_a$. Therefore, previously to combine both scores we normalize them (separately) in such a way that the maximum value of the score is equal to one, i.e. dividing by the maximum score returned. Then, the venues are sorted in decreasing order of the linear combination of the two scores, defined as

\[
CombLinear(v_j) = \lambda \times CombScore_t(v_j) + (1-\lambda) \times CombScore_a(v_j),
\]
with $\lambda$ taking its values in $[0,1]$ representing the weight given to the textual part and $1-\lambda$ the weight of the authors part.  

\section{Experiments}
\label{expe}

In this section we consider all the aspects related with the experimental evaluation of the possible merits of the proposed models, including the experimental settings (datasets, software tools, baselines and metrics being used) as well as the results obtained and their analysis.

\subsection{Experimental Settings}

\subsubsection{Dataset}
The dataset used in our experiments is a subset of the PMSC-UGR collection \cite{Albusac18}. PMSC-UGR is a collection of biomedical journal papers extracted from PubMed but also using Scopus, containing 762,508 articles. Among other information, each article includes title, abstract and keywords (full text is not available), authors’ names, which are unambiguously associated to the corresponding ORCID (Open Researcher and Contributor ID) codes, as well as the journal where the article is published (both name and ISSN)\footnote{the collection also includes citations for the majority of the articles, although we will not use this information in this work.}. We selected the journals containing at least 100 articles published between years 2007 and 2016\footnote{we did not consider one of these journals, namely PlosOne, because it had an extremely great number of articles in comparison with the other journals.}. Thus we obtained a collection having 1002 journals and 309,551 articles. We used the 276,679 articles published between 2007 and 2015 as the training set and the 33,872 articles published in 2016 as the test set. Given the size of the collection, we believe that for evaluation purposes the holdout method is appropriate and more complex methods as cross-validation are not necessary to obtain reliable results.

\subsubsection{Clustering}
In order to build the term subprofiles associated to each journal, the specific clustering algorithm that we have used is K-means \cite{Kaufman90}, more precisely its implementation within the Scikit-learn Python library\footnote{https://scikit-learn.org}. The reason for this choice is basically the efficiency of K-means, its overall good performance and the fact that our purpose is not to compare the performance of different clustering algorithms but to test our hypothesis about the suitability of using clustering for the venue recommendation problem. It should be noticed that in our case we have to deal with a document-term matrix having 276,679 rows and as many columns as the number of different terms being used.

With the aim of reducing the dimensionality, the articles in the training set were preprocessed, using also Scikit-learn. After removing stopwords and applying stemming, we also removed the terms appearing in more than 90\% of the articles (to discard terms with act as stopwords specific of this document collection) and also those terms appearing in less than 750 articles (to discard terms which are too specific)\footnote{we tried also with other values different from 750; the results were similar except in those cases where the number of remaining terms became strongly reduced, where performance degraded.}. The resulting number of terms being used is $t=4,196$.

Concerning the number of clusters $K$, we have experimented with different values, two of them obtained from classical methods in cluster analysis and another two ad hoc values:

\begin{itemize}
\item $K=\sqrt{m/2}$ \cite{Can90}, with $m$ being the number of articles in the training set (in this case $K=371$).
\item $K=m*t/e$ \cite{Kaufman90}, $t$ being the number of different terms appearing in the articles in the training set and $e$ the number of nonzero entries in the document-term matrix (in this case $e=22,694,542$ and $K=52$).
\item $K=110$, which is the number of descriptors or categories in the second level of the Medical Subject Headings (MeSH) thesaurus\footnote{https://meshb.nlm.nih.gov/treeView}.
\item $K=20$, which is the number of medical specialties described in the St. George's University\footnote{https://www.sgu.edu/blog/medical/ultimate-list-of-medical-specialties/}.
\end{itemize}

\subsubsection{Information Retrieval System} \label{sec:irs}

All the documents  have been indexed using the Lucene library\footnote{https://lucene.apache.org/}, after applying stopwords removal and stemming. Before storing an article in the index we have to consider those fields that can be used for search purposes in Lucene fields. In our case we distinguish as different fields the {\em title+abstract} (containing the textual content  of each article), {\em keywords} (comprising the keywords and subjects of the paper) and {\em authors} (we store the ORCID of each author). 

This also holds for those documents which are an aggregation of articles, as the documents $d_{jk}$ representing a topic/cluster in a journal $v_j$. Once the clustering process has finished and we know which are the articles associated to each subprofile for $v_j$ (those in $Cl_{jk}$), all these articles (the original articles, not the preprocessed versions used for clustering) are combined in a single macro-document or ``meta-article'' $d_{jk}$ representing the subprofile (joining titles-abstracts, keywords and authors in separate fields). In this case, each field contains the concatenation of all the tokens in the individual articles in $Cl_{jk}$. Note that in the case of the field {\em author} this implies that the frequency of an author in the field is the number of articles in $Cl_{jk}$ written by him.

Each article from the test set is considered (in turn) as the target article and the information associated to it is used to form a query to the IRS.  For this purpose,  we used a structured query considering the different fields stored in the index. Thus, these queries allows us to search for documents containing $w_1, w_2, \ldots$ in the {\em title+abstract} field, $s_1, s_2, \ldots $ in the {\em subject} field, in the case of the content-based model. For the author-based model the query is formed by the ORCID codes of the article's authors. Obviously, we can take advance of these fields and also use a composed query which involves both content and authorship criteria, representing a naive implementation of the combined approach (approach which will be also discussed in Section \ref{sec:results}).

The IR model that we have selected is the Language Model, specifically using the default Jelinek-Mercer smoothing \cite{Zhai01} implementation in Lucene\footnote{we have considered different IR models as BM25, vectorial, etc. and different alternatives for the required parameters. We have decided not to include this empirical study in the paper since the ones presented in the paper work properly and we do not want to enlarge excessively the experimental section.}.

\subsubsection{Baselines}
We have selected several baseline models, in order to compare their performance with that of our proposals. The first one is to use a single, monolithic profile, by aggregating all the articles in each journal into a single document and forming the corresponding document collection, as done in \cite{Silva15,Medvet14,Yang12}. We call this method {\it single profile (SP)}. The second baseline consists in considering each article published in a journal as a subprofile for this journal, i.e. using directly the original document collection (no grouping of articles is carried out), as done in \cite{Kang15,Rollins17,Luong12,Yang12,Pradhan20,Schuemie08,Errami07}. We call this method {\it distributed profile (DP)}. In both cases the processing of the corresponding document collection by the IRS is the same as previously commented, with the exception that in the case of SP it is not necessary to apply the final fusion process to get a ranking of journals. 

For comparison purposes we considered also several Machine Learning based classifiers, focusing on those  that treat text as bag of words\footnote{note that in our approach we are not considering those models that use word ordering, i.e., the semantic meaning of the document, so advanced deep learning models such as convolutional or recurrent neural networks are not taken into account.}, as SoftMax, RandomForest, Naive Bayes, Multi Layer Perceptron (MLP), Support Vector Machines, etc. In this case, the content/words of the papers represents the features whereas the venue is the target class. In this paper we will consider the two of them that have obtained the best results in a preliminary experimentation: {\it softmax}\footnote{we used the implementation available in Scikit-learn.} classifier proposed in \cite{Wang18}, which is an extension of logistic regression for solving multiclass problems and MLP\footnote{we used the implementation available in Keras, https://keras.io/}, which relies on neural networks to perfom the classification task.  

The last baseline is based on {\it doc2vec} \cite{Le14}, which has been used also as part of the system developed in \cite{Pradhan20}. Doc2vec is a technique to create alternative document representations (with dimensionality reduction) by means of vectors, which is based on shallow neural networks. In this case, we have used its Gensim Python implementation\footnote{https://pypi.org/project/gensim/}, with default parameters 
except for dimensionality of the feature vectors and the number of iterations, which were 200 and 3 respectively\footnote{we tried also other values for these two parameters. These are the values which obtained the best result.}. Once the vectors are computed in the training stage, the testing articles are also converted into vectors and a similarity measure (cosine) is computed between each vector in test and the set of training vectors. The $CombLgDCS$ fusion method is finally applied to obtain the journal rankings.

\subsubsection{Performance metrics}
In order to measure the performance of our proposals and the baselines, the common approach in the literature about venue recommendation is to match the predicted venues with the true venue where each test paper was published. Most of the works about venue recommendation use accuracy@X, i.e. the ratio between the number of correct recommendations and the number of all recommendations. A correct recommendation means that the true venue at which a test paper was published is among the first $X$ venues recommended. We have used three values of $X$, namely $X=1,5,10$. Another (also quite common) metric is mean reciprocal rank, MRR, which is the average of the inverse of the positions in the ranking at which the true venue where each test paper was published is found (0 if the true venue does not appear in the ranking)\footnote{when using this metric we only take into account the top 40 positions in the ranking.}.

\subsection{Results} \label{sec:results}

\begin{table}[tbp]
\caption{Results obtained by the Baselines}
\begin{tabular}{|c|c|c|r|r|r|r|}
\hline
\textbf{Features} & \textbf{Profile} & \textbf{Approach} & \multicolumn{1}{c|}{\textbf{acc@1}}  & \multicolumn{1}{c|}{\textbf{acc@5}} & \multicolumn{1}{c|}{\textbf{acc@10}} & \multicolumn{1}{c|}{\textbf{MRR}} \\ \hline
CB & DP & IR &{\bf 0.2282} &   {\bf 0.5370} & {\bf 0.6798}  & {\bf 0.3696} \\ \hline
CB & SP & IR & 0.2236 &   0.5278 & 0.6705 &   0.3636 \\ \hline
CB & DP & ML - MLP & 0.2012 & 0.4792 & 0.6127 & 0.3013 \\ \hline
CB & DP & ML - softmax & 0.2153 & 0.5051 & 0.6505 & 0.3504 \\ \hline
 \hline
CB & DP & ML - Doc2Vec  & 0.1545 & 0.4064 & 0.5461 & 0.2757 \\ \hline
CB & SP & ML - Doc2Vec &   0.1081 & 0.3214 & 0.4571 & 0.2145 \\ \hline
KW & SP & IR & 0.1886 &0.4436 & 0.5682 & 0.3076 \\ \hline
KW & DP & IR & 0.2027 & 0.4570 & 0.5731 & 0.3199 \\ \hline
\hline
 
AU & DP &  IR & 0.1650 & 0.4142 & 0.5107 & 0.2731 \\ \hline
AU & SP &     IR & 0.1413 & 0.3744 & 0.4897 & 0.2461 \\ 
\hline
AU & DP & ML-MLP & 0.1577  & 0.3926 & 0.5127 & 0.2412 \\ \hline
AU & DP & ML-sofmax & 0.1468   & 0.3446 & 0.4215& 0.2356 \\ \hline

\hline
\hline 
\end{tabular}
\label{tab:baseline2}
\end{table}
In order to analyze the results, two different dimensions can be considered: The first one considers those features used to represent a journal, i.e., content-based, CB (which includes text, keywords and subjects) or author-based, AU (considering only authorship). The second one is related  to the way in which the journal profile is represented, i.e., we consider a single profile, SP, a grouped profile, GP, (considering topic-based clusters)  or a distributed profile, DP (each article is considered isolately).

In Table \ref{tab:baseline2} we present the results obtained by the different baselines, including both Information Retrieval (IR) and Machine Learning (ML) approaches. From this table it can hihglighted that those methods which consider papers isolately (DP) give the best results, independently of the features (content-based or author-based) and the approach used (ML or IR). As expected, content-based features are much more useful for prediction purposes than author-based features, being able to locate the {\em right} venue within the top-10 positions 68\% of the time. We can also observe that IR-based approaches obtain better results than ML-based approaches.

The use of strategies to reduce the dimensionality of the document, as doc2vec, does not offer competitive results. For illustrative purposes, we have compared the results obtained by doc2vec with those obtained using only Keywords and Subjects (rows KW in Table \ref{tab:baseline2}), which in some way can be considered as a human-based approach to reduce the dimensionality of a document. In both cases the results are worse than those using all the content-based features.

Finally, we can highlight the results obtained using only authorship data. In this case, the 50\% of the time the correct journal is located within the top-10 recommendations. This is a clear indicator that authorship information is relevant for the task at hand, maybe because   researchers tend  to publish in a fairly small number of venues.

\begin{table}[tbp]
\caption{Results obtained using Group-based profiles }
\centering
\begin{tabular}{|c|l|r|r|r|r|}
\hline
\textbf{Features} & \textbf{Profile} &  \multicolumn{1}{c|}{\textbf{acc@1}}  & \multicolumn{1}{c|}{\textbf{acc@5}} & \multicolumn{1}{c|}{\textbf{acc@10}} &    \multicolumn{1}{c|}{\textbf{MRR}} \\ \hline
 \hline
 CB & GP, K=20 &   0.2392 & 0.5561 & 0.7014 & 0.3843 \\ \hline
 CB & GP, K=52 &  0.2428 & 0.5656 & 0.7086 & 0.3889 \\ \hline
 CB & GP, K=110 &  {\bf 0.2474} &{\bf 0.5697} & {\bf 0.7121} & {\bf 0.3928 }\\ \hline
CB & GP, K=371 &   0.2444 & 0.5631 & 0.7070 & 0.3894 \\ \hline

\hline
AU & GP, K=20 &     0.1354 & 0.3598 & 0.4758 & 0.2372 \\ \hline

 AU & GP, K=52 &     0.1389 & 0.3639 & 0.4783 & 0.2417 \\ \hline
 AU & GP, K= 110    & 0.1440 & 0.3721 & 0.4827 & 0.2477 \\ \hline

AU & GP, K=371 &     {\bf 0.1508} & {\bf 0.3812} & {\bf 0.4900} & {\bf 0.2549} \\ \hline

\hline 

\end{tabular}
\label{tab:group}
\end{table}

Now we are going to focus on our first research question: Whether considering topics to define a journal profile  is beneficial, or not, for the recommendation task. To answer this question we will focus only on the IR-based approach. The reasons are twofold: On the one hand, it gives better results in the baseline experimentation and, on the other hand, grouping the papers into topics/clusters reduces considerably the number of training data for ML approaches. The obtained results are presented in Table \ref{tab:group}, showing disparate impact in the results. On the one hand,  focusing of CB features, all the results improve the best baseline, independently of the number of topics (K) used. The best results have been obtained using $K=110$, although there is not important differences with respect to the other values of $K$. In this sense, it is clear that considering the topic helps the recommendation task. On the other hand, when considering only group-based authorship there exists a clear trend towards the use of a large set of clusters (greater values of $K$), obtaining in any case results worse than the best baseline for authors (AU+DP+IR). Note that, by using DP, this baseline can be considered as an extreme case of GP when we use the maximum number of clusters $K$, i.e., each cluster contains only one article. The clear difference in trends between the use of CB-groups with respect to the use of AU-groups might be a clear indicator that authorship represents a different source of information than content.

Finally, we will analyze in which way these two sources of information (content and authorship) can complement each other to improve the recommendations. For comparison purposes we will consider the results obtained using an IR system that benefits from the use of different fields to store content and authorship (as explained in Section \ref{sec:irs}).  Thus, in order to retrieve the relevant profiles we can use a query composed of two clauses, one related to content and the other related to authorship. This can be considered a ``naive'' approach which takes advance of the Lucene-based implementation of our IR system. The results are presented in Table \ref{tab:baselineLuc}. Comparing these results with their counterparts in Tables \ref{tab:baseline2} and \ref{tab:group}, we can conclude that combining the two types of information is beneficial for our purposes, as better results are always obtained in all the metrics. 
Nevertheless, analyzing in detail the scores and how they were computed, we realized that the content component dominates the final results. This happens because the content clause of the query usually has much more terms than the authorship one.

\begin{table}[tbp]
\caption{Results obtained for CB+AU features using queries with two Lucene-based clauses}
\centering
\begin{tabular}{|c|c|r|r|r|r|}
\hline
  \textbf{Profile} &   \multicolumn{1}{c|}{\textbf{acc@1}}  & \multicolumn{1}{c|}{\textbf{acc@5}} & \multicolumn{1}{c|}{\textbf{acc@10}} &    \multicolumn{1}{c|}{\textbf{MRR}} \\ \hline
GP, K= 20 & 0.2495 & 0.5718 & 0.7179 & 0.3963 \\ \hline
GP, K= 52 & 0.2517 & 0.5805 & 0.7228 & 0.3997 \\ \hline
GP, K= 110 & {\bf 0.2553} & {\bf 0.5830 } &\bf{ 0.7247} & \bf{ 0.4028 }\\ \hline
GP, K= 371 & 0.2521 & 0.5747 & 0.7192 & 0.3988 \\ \hline
DP & 0.2364 & 0.5489 & 0.6901 & 0.3764 \\ \hline
SP & 0.2342 & 0.5463 & 0.6896 & 0.3776 \\ \hline

\end{tabular}
\label{tab:baselineLuc}
\end{table}

\begin{figure}
\caption{acc@10 obtained when using CB+AU features in an IR approach using different profiles and lambda ($\lambda$) values for the linear combination. } \label{fig}
\centering
 \includegraphics[width=0.9\textwidth]{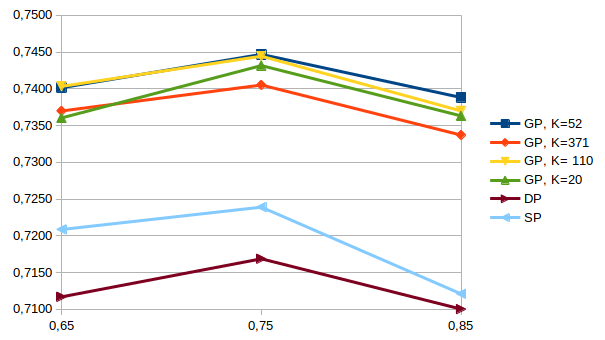}
\end{figure}

The linear combination proposed in  Section \ref{combination} can help to  tackle this problem by normalizing the  output scores for each component. By means of this   combination we can  control explicitly how each component contributes to the final result. Particularly in Table \ref{CB+AU} and Figure \ref{fig} we illustrate the performance of this approach for different content-based profiles, fixing AU+DP+IR as the authorship component\footnote{in a preliminary experimentation other combinations have been tested, but we obtained the best results with this one.} and varying the parameter $\lambda$, which controls the weight given to the content-based component. 

From Table \ref{CB+AU} we can confirm that the use of topic-based profiles (GP) for the content component and the use of the linear combination is highly recommended to perform this task, since we always get better values than their counterparts obtained using a ``naive'' approach (Table \ref{tab:baselineLuc}). With respect to the number of topics used we could choose K=110, i.e., the number of categories in the second level of MeSH thesaurus. However, the results obtained using K=52, i.e., $K=m*t/e$, are similar, so we recommend the use of this alternative when there is no such domain information. Nevertheless, there are no big differences among the results obtained with different $K$, which might imply that its particular value is not a critical parameter for tuning purposes. With respect to the value of $\lambda$, the use of values in the range $0.75$ seems to be a good alternative, representing improvements with respect the best baseline of 16.4\%, 12.6\%, 13.8\% and 12.9\%  in acc@1, acc@5, acc@10 and MRR, respectively. Although not presented in the results, we have experimentaly proved that performance declines with values of $\lambda$ outside the range 0.65--0.85.

\begin{table}[htbp]
\caption{Results obtained using the linear combination of CB and AU}
\centering 
\begin{tabular}{|l|r|r|r|r|r|}
\hline
\textbf{Profile} & \multicolumn{1}{c|}{\textbf{$\lambda$}} & \multicolumn{1}{l|}{\textbf{acc@1}} & \multicolumn{1}{l|}{\textbf{acc @5}} & \multicolumn{1}{l|}{\textbf{acc@10}} & \multicolumn{1}{l|}{\textbf{MRR}} \\ \hline
GP, K=20 & 0.65 & 0.2637 & 0.5962 & 0.7360 & 0.4139 \\ \hline
GP, K=20 &  0.75  & 0.2655 & 0.6032 & 0.7432 & 0.4169 \\ \hline
GP, K=20 & 0.85 & 0.2555 & 0.5913 & 0.7363 & 0.4075 \\ \hline
GP, K=52 & 0.65 & 0.2652 & 0.5986 & 0.7401 & 0.4159 \\ \hline
GP, K=52 &  0.75  & 0.2656 & 0.6045 & \textbf{0.7447} & 0.4174 \\ \hline
GP, K=52 & 0.85 & 0.2597 & 0.5958 & 0.7388 & 0.4105 \\ \hline
GP, K= 110 &  {0.65} & \textbf{0.2670} & 0.6036 & 0.7403 & \textbf{0.4177} \\ \hline
GP, K= 110 & 0.75 & 0.2652 & \textbf{0.6070} & 0.7444 & \textbf{0.4177} \\ \hline
GP, K= 110 & 0.85 & 0.2596 & 0.5975 & 0.7370 & 0.4107 \\ \hline
GP, K=371 & 0.65 & 0.2637 & 0.5987 & 0.7370 & 0.4141 \\ \hline
GP, K=371 & 0.75 & 0.2630 & 0.6004 & 0.7405 & 0.4141 \\ \hline
GP, K=371 & 0.85 & 0.2562 & 0.5889 & 0.7337 & 0.4063 \\ \hline

DP &  0.65 & 0.2525 & 0.5733 & 0.7117 & 0.3979 \\ \hline
DP & 0.75 & 0.2452 & 0.5739 & 0.7169 & 0.3935 \\ \hline
DP & 0.85 & 0.2369 & 0.5627 & 0.7100 & 0.3849 \\ \hline
SP & 0.65 & 0.2461 & 0.5766 & 0.7208 & 0.3962 \\ \hline
SP &  {0.75} & 0.2554 & 0.5835 & 0.7239 & 0.4043 \\ \hline
SP & 0.85 & 0.2496 & 0.5713 & 0.7121 & 0.3959 \\ \hline

\end{tabular}
\label{CB+AU}
\end{table}

\section{Concluding Remarks}
\label{conclu}
In this paper we tackle the publication venue recommendation problem. The main purpose is to identify an appropriate journal or conference where a research paper could be submitted. In our approach we  used the content of the papers which had been published in each venue in order to define the venue subprofiles. These subprofiles have been generated using a document clustering algorithm based on the textual content of the papers, in such a way that related papers are joined under the same subprofile. Then, the recommendations were obtained by matching the target paper and the venues subprofiles through an information retrieval system.

Our experimentation demonstrates that the use of topic-based clusters of papers helps to improve the recommendations consistently, compared with several baseline methods, either based on Information retrieval but not using topic subprofiles, or based on machine learning classifiers. Our hypothesis is that this happens because the different subprofiles define better and more precisely the scope of each venue. We have also studied the role of authorship in the recommendation problem showing that it is an important and complementary piece of information. The use of a linear combination between the normalized rankings obtained by the content-based and the authorship-based approaches allows us to obtain the best results. 

As future work, it can be interesting to study different alternatives to learn the topic-based clusters, as it can be the use of Latent Dirichlet Allocation and study how these clusters can be used to explain the recommendations to the final users. Also, we plan to extend our system to include the use of citations and references (both the text associated to them and the resulting network) and the inclusion of more sophisticated author networks which can represent the relationships between authors.
Finally, with respect to the machine learning approach, although in our experimentation we do not get good results, we believe that the use of more sophisticated deep learning models could improve the results.

\subsection*{Acknowledgements}
This work was supported by the Spanish ``Agencia Estatal de Investigaci\'on'' [grant number PID2019-106758GB-C31/AEI/10.13039/501100011033]; 
the Spanish ``Programa operativo FEDER Andalucía 2014-2020 de la Junta de Andalucía y la Universidad de Granada'' [grant number A‐TIC‐146‐UGR20]; and the European Regional Development Fund (ERDF-FEDER).

\end{document}